\documentclass[authoryear,preprint,review,12pt]{elsarticle}

\newcommand{\Msun}{\mbox{$M_{\odot}$}}
\newcommand{\Rsun}{\mbox{$R_{\odot}$}}
\newcommand{\Lsun}{\mbox{$L_{\odot}$}}
\newcommand{\kms}{\mbox{km s$^{-1}$}}

\usepackage{graphics}

\journal{New Astronomy}

\begin{document}

\begin{frontmatter}



\title{Photometric and spectroscopic observations of the F3+M3 eclipsing binary T-Lyr0-08070}


\author[OCakirli]{\"{O}m\"{u}r \c{C}ak{\i}rl{\i}\corref{cor1}}
\ead{omur.cakirli@gmail.com}
\cortext[cor1]{Corresponding author}

\address[OCakirli]{Ege University, Science Faculty, Astronomy and Space Sciences Dept., 35100 Bornova, \.{I}zmir, Turkey.}
\author[OCakirli]{Cafer Ibanoglu} 
\author[OCakirli]{E. Sipahi}

\begin{abstract}
The multi-color photometric and spectroscopic observations of the newly discovered eclipsing binary T-Lyr0-08070 were obtained. The 
resultant light and radial velocities were analysed and the absolute parameters of the components were determined. The system is 
composed of an F3 and an M3 main-sequence stars. Masses and radii were estimated to be 1.37$\pm$0.23 M$_{\odot}$ and 
1.60$\pm$0.09 R$_{\odot}$ for the primary and 0.32$\pm$0.04 M$_{\odot}$ and 0.86$\pm$0.06R$_{\odot}$ for the secondary 
star. The less massive secondary component has a radius at least two times larger with respect to its mass. Using the BVJHK 
magnitudes of the system we estimated an interstellar reddening of 0.22 mag and a  distance to the system as 479$\pm$36 pc.    
\end{abstract}

\end{frontmatter}


\section{Introduction}\label{S1}
Transit events for the T-Lyr0-08070 (V$_{max}$=12.15 mag.) were detected during regular operations of the Trans-Atlantic Exoplanet 
Survey network ({\em TrES}). \citet{Fernandez_2009} classified this star as a long-period single-lined eclipsing binary with an F-star primary 
and an {\em unseen} M-dwarf secondary. They obtained follow-up spectroscopic observations using 1.5\,m Wyeth Reflector at the Oak Ridge 
Observatory and 1.5\,m Tilinghast Reflector at Whipple Observatory. In order to reveal fundamental characteristics of the T-Lyr0-08070 they 
combined photometric data from {\em TrES} survey and {\em KeplerCam} photometry with those single-order echelle spectra in a 
wavelength window of 45 \AA~ centered at 5187 \AA.~ They obtained the radial velocities only for the F-star. Since their spectra 
cover a narrow wavelength interval, there is a clear dependence between the effective temperature and the metal abundance of 
the star. For this reason they have, modeled the spectra for metallicity index of -1.0, -0.5, 0.0 and +0.5 to obtain effective 
temperature, projected rotational velocity and surface gravity of the primary star. \citet{Fernandez_2009} modeled the 
{\em KeplerCam} primary transit light curve employing the method of \citet{Mandel_Agol}, which was an 
analytical light curve analysis developed for planetary transit searches. Assuming orbit-rotation synchronization and
adopting Fe/H=-0.5, the mass and radius of the primary star were derived as 0.95 \Msun and 1.36 \Rsun. Under the same assumptions, they
derived the mass and radius of the secondary star as 0.345 \Msun and 0.360 \Rsun.

We planned new spectroscopic and photometric observations of the T-Lyr0-08070 to obtain the masses, radii and effective 
temperatures of the primary and secondary stars. The photometric light curves were obtained in the  BVRI passbands and 
analysis are presented. The radial velocities for both components are also obtained and analysed.

\section{Data acquisition}                                                                                                             \label{sec:obs}
\subsection{Spectroscopy}
Optical spectroscopic observations of the T-Lyr0-08070 were obtained with the Turkish Faint Object Spectrograph Camera 
(TFOSC)\footnote{http://tug.tug.tubitak.gov.tr/rtt150\_tfosc.php} attached to the 1.5 m telescope on 8 nights (July, 2010) 
under good seeing conditions. Further details on the telescope and the spectrograph can be found at http://www.tug.tubitak.gov.tr. The 
wavelength coverage of each spectrum was 4000-9000 \AA~in 12 orders, with a resolving power of $\lambda$/$\Delta \lambda$ 
7\,000 at 6563 \AA~and an average signal-to-noise ratio (S/N) was $\sim$120. The exposure times were between 10 and 60 minutes during 
observations. We also obtained a high S/N spectrum of the M dwarf GJ\,740 
(M0 V) and GJ\,623 (M1.5 V) for use as templates in derivation of the radial velocities \citep{Nidever}. 

The electronic bias was removed from each image and we used the 'crreject' option for cosmic ray removal. Thus, the resulting 
spectra were largely cleaned from the cosmic rays. The echelle spectra were extracted and wavelength calibrated by using Fe-Ar 
lamp source with help of the IRAF {\sc echelle} package \citep{Tonry_Davis}. 
 
The stability of the instrument was checked by cross correlating the spectra of the standard star against each other using 
the {\sc fxcor} task in IRAF. The standard deviation of the differences between the velocities measured using {\sc fxcor} and the 
velocities in \citet{Nidever}  was about 1.1 \kms.

\subsubsection{Spectral classification}
We have used our spectra to reveal the spectral type of the primary component of T-Lyr0-08070. For this purpose we have degraded the 
spectral resolution from 7\,000 to 3\,000, by convolving them with a Gaussian kernel of the appropriate width, and we have measured
the equivalent widths ($EW$) of photospheric absorption lines for the spectral classification. We have followed the procedures of 
\citet{hernandez}, choosing helium lines in the blue-wavelength region, where the contribution of the secondary component to the 
observed spectrum is almost negligible. From several spectra we measured $EW_{\rm He I+ Fe I\lambda 4922 }=0.38\pm 0.02$\,\AA~and 
$EW_{\rm H_\beta 4861 }=5.05\pm 0.11$\,\AA. From the calibration relations $EW$--Spectral-type of \citet{hernandez}, we have 
derived a spectral type of F3$\pm 2$ for the primary star. In Fig.1 we compare the spectrum of T-Lyr0-08070 with the spectra of 
standard stars HD 224639 (F0V), HD 223623 (F2V) and HD 18215 (F5V). The spectra of standard stars are taken from the ELODIE archive\footnote{http://atlas.obs-hp.fr/elodie/}
database. This comparison confirms the spectral-type of the more massive star derived from the EW measurements.
  
\begin{figure*}
\includegraphics[width=10cm]{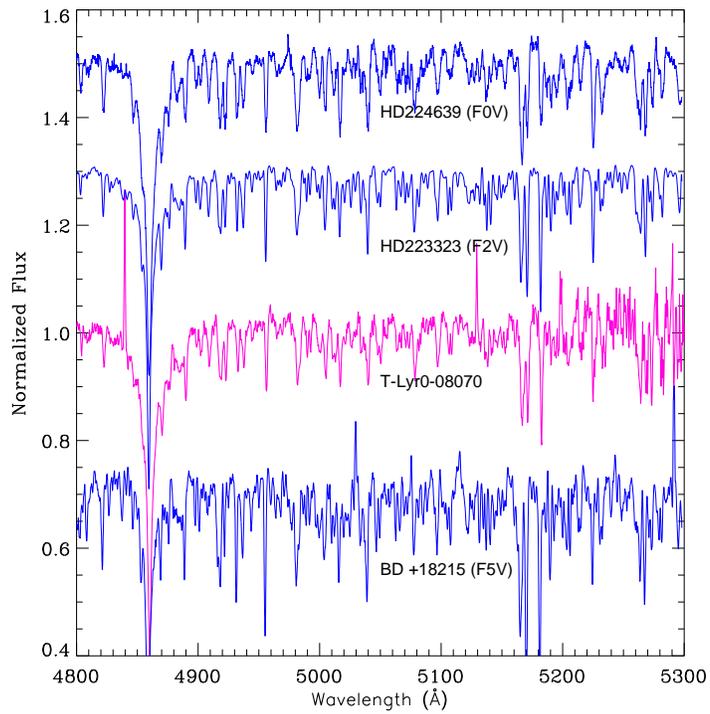}
\caption{Comparison of the observed spectrum of the binary with those of the F(0-5)V spectral type standards in the $\lambda\lambda 4800-5300$\,\AA~region.} \end{figure*}

\subsubsection{Rotational velocity}
The width of the cross-correlation profile is a good tool for the measurement of $v \sin i$ (see, e.g., Queloz et al. 1998). The 
rotational velocities ($v \sin i$) of the two components were obtained by measuring the FWHM of the CCF peaks in nine high-S/N spectra 
of T-Lyr0-08070 acquired close to the quadratures, where the spectral lines have the largest Doppler-shifts. In order to construct a calibration 
curve FWHM--$v \sin i$, we have used an average spectrum of the HD\,128429, acquired with the same instrumentation. Since the rotational 
velocity of HD\,128429 is very low but not zero  (F5\,V, $v \sin i$ $\simeq$14 km s$^{-1}$, e.g., \citet{Royer02}), it could be 
considered as a useful template for F-type stars rotating faster than $v \sin i$ $\simeq$ 10 km s$^{-1}$. The spectrum 
of HD\,128429 was synthetically broadened by convolution with rotational profiles of increasing $v \sin i$ in steps of 5 km s$^{-1}$ 
and the cross-correlation with the original one was performed at each step. The FWHM of the CCF peak was measured and the 
FWHM-$v \sin i$ calibration was established. The $v \sin i$ values of the components of T-Lyr0-08070 were derived from the 
FWHM of their CCF peaks and the aforementioned calibration relations, for a few wavelength regions for the best spectra. This 
gave values of 58$\pm$1 km s$^{-1}$ for the primary and 25$\pm$9 km s$^{-1}$ for the secondary star. The rotational velocity of 
the primary star is in a good agreement with that estimated by \citet{hernandez}. 

\begin{figure}
\includegraphics[width=8cm]{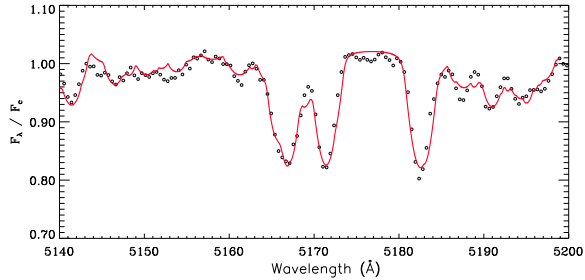}
\caption{Observed spectrum (points) of T-Lyr0-08070 in the vicinity of Mg\,{\sc I} triplet. The synthetic spectrum
(F3\,V+M3\,V) is displayed with continuous line.} \end{figure}

For construction of the synthetic spectrum of the system the spectra of the standard 
stars ($\iota$ Psc and GJ\,740), obtained with the same  instrumentation, have been rotationally broadened by convolution with the appropriate rotational 
profile and then co-added, properly weighted by using physical parameters ($T_1$, $T_2$, $R_1$, $R_2$, $vsini_{12}$) of the components 
as input parameters and, then, Doppler-shifted according to the radial velocities of the components. 
$T_1$ and $vsini_{1,2}$ were determined in \S2.1.1 and \S2.1.2. $T_2$ and $R_{1,2}$ were obtained from the combination of 
the light and radial velocity curves analyses (see \S4). In Fig.\,2 we compare the observed spectrum of T-Lyr0-08070 around the Mg\,{\sc I} triplet 
with the synthetic one for a system consisting of F3V+M3V type stars.

\subsection{Photometric observations}
In order to provide high-quality light curves of the system in different passbands, we used {\sf SI\,1100 CCD 
Camera\footnote{http://www.specinst.com/}} mounted on the 1\,m telescope at the T\"{U}B{\.I}TAK National Observatory, Turkey. 
SI\,1100 Camera utilizes a {\em Cryogenic} cooler 4k$\times$4k, Fairchild 486 CCD that gives a 21$^{\prime}\times21^{\prime}$ 
field and a pixel size of 0$^{\prime\prime}$.72 when the binning is 2$\times$2. Differential aperture photometry was performed 
to obtain the light curves. The observed magnitudes of T-Lyr0-08070 in the B and V passbands were compared with those
for the comparison and check stars. The coordinates and magnitudes of the stars observed are given in Table\,1. The B and V 
magnitudes of the variable are obtained in this study. The typical standard deviations of the differential magnitudes are about 0.005 mag.

\begin{figure}
\includegraphics[width=9cm,angle=0]{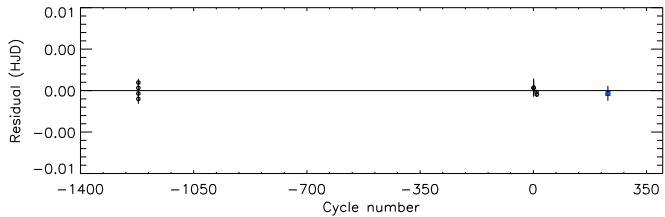}
\caption{\label{fig:periodresids} The residuals between the observed times of minimum light and computed with the new ephemeris. The square refers 
to the secondary minimum.} \end{figure}

\begin{table} 
\footnotesize
\caption{ The coordinates and BVJHK magnitudes of the stars observed.}      
  \setlength{\tabcolsep}{1.2pt}
\begin{tabular}{lccccccc} \hline
Star	& $\alpha$&$\delta$    &B (mag)	&V (mag)&J (mag)&H (mag)&K (mag)	\\
\hline					
T-Lyr0-08070	    & 19$^h$ 19$^m$ 04$^s$	& 38$^{\circ}$ 40$^\prime$ 57$^{\prime\prime}$     & 12.747 	& 12.150 	& 11.507		& 11.321	& 11.261		\\
GSC 3121-01011    	& 19$^h$ 19$^m$ 50$^s$	& 38$^{\circ}$ 40$^\prime$ 57$^{\prime\prime}$     & 10.81  	& 9.92   	& 8.173  		& 7.694	& 7.633		\\	
GSC 3121-00379    	& 19$^h$ 19$^m$ 33	$^s$	& 38$^{\circ}$ 40$^\prime$ 57$^{\prime\prime}$     & 11.7   	& 11.5   	& 10.703 		& 10.569	& 10.544		\\
\hline
\end{tabular}
\end{table}

\subsection{Effective temperature of the primary star}
We find an observed color of B-V=0.60$\pm$0.02 mag for the binary system T-Lyr0-08070 at out-of-eclipse which should be
reddened if the spectral type F3 is considered. The observed infrared colors of J-H=0.186$\pm$0.039 and H-K=0.060$\pm$0.035 given in 
the 2MASS catalog \citep{cutri} correspond to a spectral type of F3$\pm$2 is in a good agreement with that we derived by spectroscopy. 

The effective temperature deduced from the calibrations of spectral type-effective temperature or intrinsic B-V color-effective 
temperature given by \citet{drill}, \citet{dejager}, \citet{Alonso}, \citet{flower} and \citet{popper} is about 6\,750$\pm 150$\,K. However 
we estimated a temperature of 6\,670$\pm$170\,K for the primary star from the infrared colors - effective temperature calibrations 
of \citet{tokunaga}. Temperature uncertainty of the primary component results from considerations of spectral type uncertainties, and 
calibration differences. As a weighted mean we adopted 6\,700$\pm$150 K for the effective temperature of the primary star.

\begin{figure}
\includegraphics[width=9cm=0]{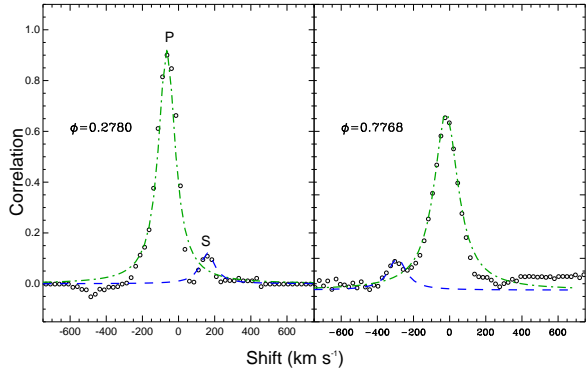}
\caption{Sample of CCFs between T-Lyr0-08070 and the RV template spectrum around the first and second quadrature.} 
\end{figure}

\section{Period determination}                                   
We obtained five times of minimum light for the system during photometric observations. These times of minima are given in 
Table\,2 as averaged for the filters used. The times of minimum light obtained by \citet{Fernandez_2009} were also included in the 
list. A linear least-squares fit to the data yields the following ephemeris  \[ {\rm Min\,I} = {\rm HJD}\ 2455443.3232 (2) + 1^d.18483992 (41) \times E \] 
where the bracketed quantity is the uncertainty in the last digit(s) of the preceding number. All uncertainties quoted in this work 
are standard errors. The residuals of the fit are plotted in Fig.\,3 and show no indication of any form of period change in about five years.

\begin{table} \begin{center}
\caption{Times of minimum light for T-Lyr0-08070 and the residuals between the observed
and calculated times with the new ephemeris.
\newline {\bf References:} (1) \citet{Fernandez_2009}; (2) This work.}
\begin{tabular}{cccc} \hline
{\em Time of minimum} & {\em Cycle number} &$O-C$&{\em Reference}  \\
{\em (HJD 2\,500\,000+)} &&&  \\
\hline
3996.6326$\pm$0.0001	&	-1221.0	&	-0.0010	&	1					\\
3996.6333$\pm$0.0001	&	-1221.0	&	-0.0004	&	1					\\
3996.6340$\pm$0.0001	&	-1221.0	&	 0.0003	&	1					\\
3996.6346$\pm$0.0001	&	-1221.0	&	 0.0009	&	1					\\
5443.3231$\pm$0.0004	&	 0.0			&	 0.0003	&	2					\\
5444.5083$\pm$0.0002	&	 1.0			&	 0.0002	&	2					\\
5455.1708$\pm$0.0004	&	 10.0		&	-0.0008	&	2					\\
5456.3550$\pm$0.0003	&	 11.0		&	-0.0014	&	2					\\
5716.4286$\pm$0.0003  &	230.5	    & 	-0.0002    & 2 					\\
\hline \end{tabular} \end{center} \end{table}

\section{Analysis}																		\label{sec:analysis}
\subsection{Radial velocity curve}
To derive the radial velocities for the components of binary system, the 12 TFOSC spectra of the eclipsing binary  were cross-correlated against 
the spectrum of GJ\,740, a single-lined M0V star, on an order-by-order basis using the {\sc fxcor} package in IRAF. The majority of the spectra 
showed two distinct cross-correlation peaks in the quadrature, one for each component of the binary. Thus, both peaks were fit independently
in the quadrature with a $Gaussian$ profile to measure the velocity and errors of the individual components. If the two peaks appear 
blended, a double Gaussian was applied to the combined profile using {\it de-blend} function in the task. For each of the 12 observations 
we determined a weighted-average radial velocity for each star from all orders without significant contamination by telluric absorption
features. Here we used as weights the inverse of the variance of the radial velocity measurements in each order, as reported 
by {\sc fxcor}.

We adopted a $two-Gaussian$ fit algorithm to resolve cross-correlation peaks near the first and second quadratures when spectral lines are 
visible separately. Fig.\,4 shows examples of cros-correlations obtained by using the largest FWHM at nearly first and second quadratures. The stronger 
peaks in each CCF correspond to the more luminous component which has a larger weight into the observed spectrum. 

\begin{table}
\centering
\begin{minipage}{85mm}
\caption{Heliocentric radial velocities of T-Lyr0-08070. The columns give the heliocentric Julian date, the
orbital phase (according to the ephemeris in Eq.~1), the radial velocities of the two components with the 
corresponding standard deviations.}

\begin{tabular}{@{}ccccccccc@{}c}
\hline
HJD 2400000+ & Phase & \multicolumn{2}{c}{Star 1 }& \multicolumn{2}{c}{Star 2 } 	\\
             &       & $V_p$                      & $\sigma$                    & $V_s$   	& $\sigma$	\\
\hline
55390.3850  &0.3204 &-70.5 &-3.5 	&--    		&--		\\
55391.4725  &0.2382 &-72.2 &-3.6 	&-- 			&--		\\
55392.3905  &0.0130 &-34.3 &11.1 	&--    		&--		\\
55392.5751  &0.1688 &-73.0 &-3.7 	&--    		&--		\\
55393.2954  &0.7768 & 13.8 & 0.7 	&-210.7	&9.9		\\
55393.3438  &0.8176 &  9.6  & 0.5 	&-199.4	&11.1	\\
55394.5068  &0.7992 & 12.9 & 0.6 	&-207.9	&12.8	\\
55396.3351  &0.3422 &-66.3 &-3.3 	&--    		&--		\\
55396.4813  &0.4656 &-40.2 & 7.4 	&--    		&--		\\
55397.4438  &0.2780 &-73.9 &-3.7 	&151.5 	&10.4	\\
55398.5238  &0.1894 &-74.7 & 6.2 	&--    		&--		\\
55398.5700  &0.2285 &-76.4 &-3.8 	&149.9 	&10.4	\\
\hline \\
\end{tabular}
\end{minipage}
\end{table}

The heliocentric RVs for the primary (V$_p$) and the secondary (V$_s$) components are listed in Table\,3 and plotted in Fig.\,5 
against the orbital phase.  The RVs listed in Table\,3 are the weighted averages of the values obtained from the cross-correlation 
of orders \#4, \#5, \#6 and \#7 of the target spectra with the corresponding order of the standard star spectrum. The weight 
$W_i = 1/\sigma_i^2$ has been given to each measurement. The standard errors of the weighted means have been calculated on 
the basis of the errors ($\sigma_i$) in the RV values for each order according to the usual formula (e.g.\citet{toping}). The $\sigma_i$ 
values are computed by {\sc fxcor} according to the fitted peak height, as described by \citet{Tonry_Davis}.

\begin{figure}
\includegraphics[width=9cm,angle=0]{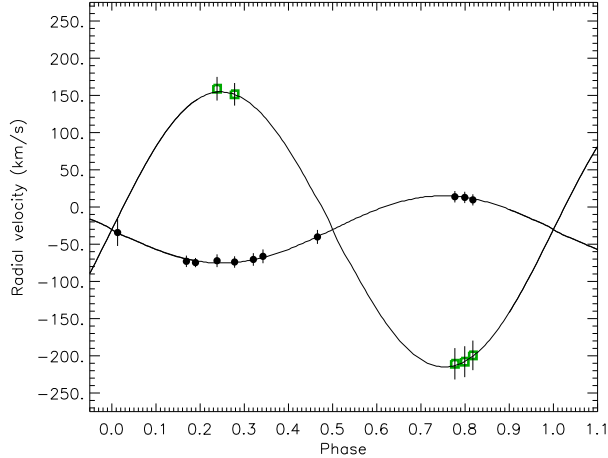}\\
\caption{Radial velocity curve folded on a period of 1.1848 days. Points with error bars (error bars are masked by 
the symbol size in some cases) show the radial velocity measurements for the components of the system 
(primary: filled circles, secondary: open squares). } \end{figure}

First we analysed the radial velocities for the initial orbital parameters. We used the orbital period held fixed 
and computed the eccentricity of the orbit, systemic velocity and semi-amplitudes of the RVs. The results of the analysis 
are as follow: $\gamma$= -30$\pm$2 \kms, $K_1$=45$\pm$3 and $K_2$=190$\pm$12 \kms with a circular orbit. Using 
these values we estimate the projected orbital semi-major axis and mass ratio as: $a$sin$i$=5.50$\pm$0.06 \Rsun~ and 
$q=\frac{M_2}{M_1}$=0.237$\pm$0.008.

\begin{figure*}
\center
\includegraphics[width=10cm,angle=90]{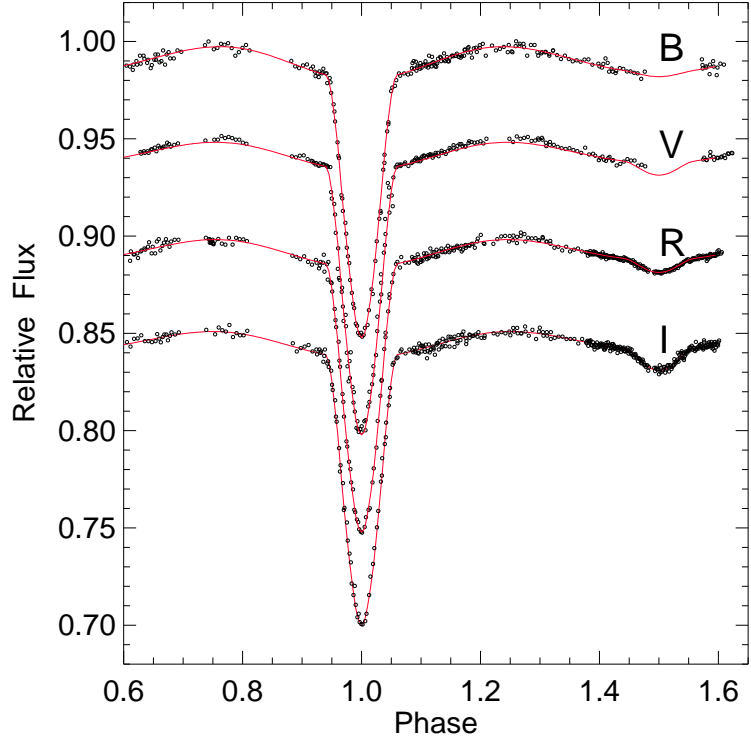}
\caption{The observed and computed light curves of T-Lyr0-08070. } \end{figure*}

\subsection{Light curve modeling}
We used the most recent version of the eclipsing binary light curve modeling algorithm of \citet{wil} (with updates), as 
implemented in the {\sc phoebe} code of Pr{\v s}a \& Zwitter (2005). The code needs some input parameters, which 
depend upon the physical properties of the component stars. In the light curve solution we fixed some parameters whose 
values can be estimated from global stellar properties, such as effective temperature of the hotter component and mass-ratio 
of the system. The effective temperature of the primary star has already been derived from photometry and spectroscopy as 
6\,700 K and the mass-ratio as 0.237. A preliminary estimate for the effective temperature of the cooler component is made 
comparing the depths of the eclipses and mass-ratio. We adopted the linear limb-darkening coefficients from Van Hamme 
(1993) for the primary and secondary components, and the bolometric albedos from \citet{lucy}, for each passband. The gravity 
brightening coefficients are taken as 0.32 for both components. The rotational velocities of the components are assumed to 
be synchronous with the orbital one. The adjustable parameters in the light curves fitting were the orbital inclination, the 
surface potentials, effective temperature of the secondary and the luminosity of the primary. Our final results of simultaneous 
solution are listed in Table\,4 and the computed light curves (continuous line) are compared with the observations in 
Fig.\,6. The fractional radii ($r_{1,2}$) given in Table\,4 are the mean values. The proximity effects at out-off-eclipses 
are clearly seen. The uncertainities assigned to the adjusted parameters are the internal errors provided directly by the 
Wilson-Devinney code.

\begin{table}
\scriptsize
\caption{Results of the simultaneous B, V, R, and I passbands light curve analysis for T-Lyr0-08070.}
\begin{tabular}{lr}
\hline
Parameters & Adopted  \\
\hline	
$i^{o}$			               			 						&78.54$\pm$0.51\\
T$_{eff_1}$ (K)												&6\,700[Fix]\\
T$_{eff_2}$ (K)												&3\,200$\pm$180\\
$\Omega_1$													&3.650$\pm$0.012\\
$\Omega_2$													&3.314$\pm$0.089\\
$r_1$																&0.285$\pm$0.004\\
$r_2$																&0.154$\pm$0.009\\
$\frac{L_{1}}{(L_{1}+L_{2})}$ (B) 				&0.998$\pm$0.008\\
$\frac{L_{1}}{(L_{1}+L_{2})}$ (V) 				&0.988$\pm$0.008\\
$\frac{L_{1}}{(L_{1}+L_{2})}$ (R) 				&0.910$\pm$0.008\\
$\frac{L_{1}}{(L_{1}+L_{2})}$ (I) 				&0.888$\pm$0.008\\
$\chi^2$															&0.006\\				
\hline
\end{tabular}
\end{table}

\section{Absolute dimensions and distance to the system}
The semi-amplitude of the radial velocities are obtained from the radial velocity analysis. The inclination of the orbit and the 
fractional radii of the components are determined from the light curve solutions. Therefore, we can compute the absolute parameters of the 
component stars. We present the fundamental physical parameters for the system T-Lyr0-08070 with their uncertainties in Table\,5. The 
absolute parameters we obtained in this study are significantly different from those obtained by \citet{Fernandez_2009}, because they derived 
the parameters for only the transit light curve and the rotational velocity of the primary star. In particular, the mass and radius of the secondary 
star are very different when compared with those of \citet{Fernandez_2009}. 
They estimated a mass for the less massive star as 0.22, 0.25, 0.29 and 0.32 \Msun~, and radius as 0.25, 0.27, 0.29 and 0.31 \Rsun from the 
stellar isochrones for the values of Fe/H=-1.0, -0.5, 0.0 and +0.5, respectively. Using orbit-rotation synchronization they find mass and radius for 
the secondary star as 0.240 \Msun and 0.265 \Rsun. Our radial velocity and light curve analyses yield a mass of 0.32 and a radius of 0.86 in solar units.

In Fig.\,7 we compare locations of the low-mass stars in the mass-radius diagram. The radius of the M3 dwarf is at least two times larger 
than that predicted from the stellar models. The less massive components of the close binary systems having a mass smaller than the sun 
are  generally oversized with respect to the theoretical models as indicated recently by \citet{cakirli}. However, their effective temperatures 
are generally lower than those of single main sequence stars with similar mass. Larger radii and lower effective temperatures are 
attributed to the solar-like activities (see Morales, Ribas \& Jordi (2008)). Large  spots on the  stars prevent energy output, therefore, the 
star expands. While the radius increases the surface temperature decreases in active stars in close binary systems.

The BVJHK magnitudes of the binary and the comparison stars given in Table\,1 allow us estimation of the color excess. Assuming the spectral 
type of the primary star as F3 we estimated the interstellar reddening as E(B-V)=0.22$\pm0.02$ mag 
using the tables given by \citet{drill}. Using the radius and effective temperature of the primary star we compute its luminosity as 4.56 \Lsun. Then taking 
this value and bolometric absolute magnitude for the sun as 4.74 one may compute its absolute bolometric magnitude as 3.08 "mag". The 
bolometric correction of -0.12 "mag" taken from  \citet{drill} we find an absolute visual magnitude of 3.21 "mag" 
for the more massive star. The apparent visual magnitude of 12.162 extracting the light contribution of the secondary star to the total 
light, the absolute visual magnitude for the primary star and interstellar extinction of A$_V$=0.68 "mag" yield a distance to the system 
as 479$\pm$36 pc. However, JHK magnitudes lead to a distance of about 630 pc with larger standard deviation. This difference may be originated 
from the bolometric corrections taken from different tables (Girardi et al. (2002) and Kervella et al. (2004)).

\begin{table}
 \setlength{\tabcolsep}{2.5pt} 
  \caption{Fundamental parameters of T-Lyr0-08070.}
  \label{parameters}
  \begin{tabular}{lcc}
  \hline
   Parameter 																& Primary	&	Secondary				\\
   \hline
   Spectral Type															& F3$\pm$2  	& M3$\pm$1    			\\
   $a$ (R$_{\odot}$)														&\multicolumn{2}{c}{5.61$\pm$0.06}		\\
   $V_{\gamma}$ (km s$^{-1}$)												&\multicolumn{2}{c}{-30$\pm$3}			\\
   $i$ ($^{\circ}$)															&\multicolumn{2}{c}{78.5$\pm$0.5}		\\
   $q$																		&\multicolumn{2}{c}{0.237$\pm$0.008}	\\
   Mass (M$_{\odot}$) 														& 1.37$\pm$0.23 & 0.32$\pm$0.04			\\
   Radius (R$_{\odot}$) 													& 1.60$\pm$0.09 & 0.86$\pm$0.06			\\
   $\log~g$ ($cgs$) 														& 4.17$\pm$0.03 & 4.08$\pm$0.05			\\
   $T_{eff}$ (K)															& 6700$\pm$150	& 3170$\pm$150			\\
   $(vsin~i)_{obs}$ (km s$^{-1}$)											& 58$\pm$1		& 25$\pm$9				\\
   $(vsin~i)_{calc.}$ (km s$^{-1}$)											& 69$\pm$2		& 27$\pm$2				\\
   $\log~(L/L_{\odot})$														& 0.67$\pm$0.06	& -1.17$\pm$0.10		\\
   $d$ (pc)																	& \multicolumn{2}{c}{479$\pm$36}		\\
\hline  
  \end{tabular}
\end{table}

\begin{figure}
\includegraphics[width=9cm]{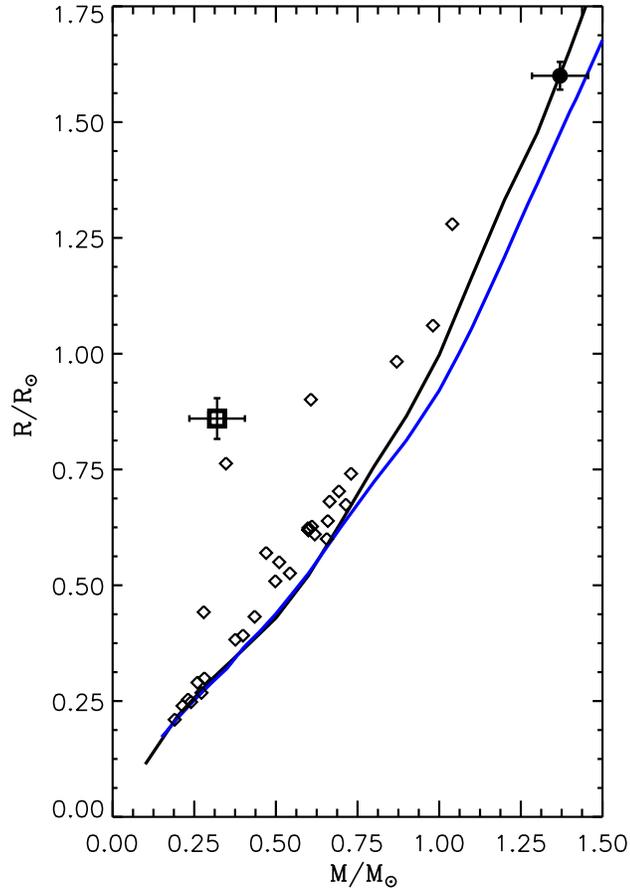}
\caption{\label{fig:periodresids} Locations of the low-mass stars in the mass-radius diagram. The less massive companion of the 
T-Lyr0-08070 is shown by a large square and the primary by dot with standard deviations.} \end{figure}

\section*{Acknowledgments}
We thank T\"{U}B{\.I}TAK National Observatory (TUG) for a partial support in using RTT150 and T100 telescopes with project numbers 
10ARTT150-483-0, 11ARTT150-123-0 and 10CT100-101. We also thank responsible for the Bak{\i}rl{\i}tepe observing station for their 
warm hospitality. This research has been made use of the ADS and CDS databases, operated at the CDS, Strasbourg, France and T\"{U}B\.{I}TAK 
ULAKB{\.I}M S\"{u}reli Yay{\i}nlar Katalo\v{g}u. The authors are grateful to the anonymous referee for his/her valuable comments.

\end{document}